\documentclass[preprint,amsmath]{revtex4-1}
\usepackage{amsmath}
\usepackage{amssymb}
\usepackage{color}
\usepackage[normalem]{ulem}
\usepackage{mathtools}
\usepackage{color}

\begin{document}
\title{ Packets of Diffusing Particles Exhibit Universal Exponential Tails }
\author{Eli Barkai}
\author{Stanislav Burov}
\email{stasbur@gmail.com}

\affiliation{Physics Department, Bar-Ilan University, Ramat Gan 5290002,
Israel}

\pacs{PACS}

\begin{abstract}
 Brownian motion is a Gaussian process described by the central limit theorem. However, exponential decays of the positional probability density function $P(X,t)$ of packets of spreading random walkers, were observed in numerous  situations that include glasses, live cells and bacteria suspensions. 
 We show that such exponential behavior is generally valid in a large class of problems of transport in random media.
 %This fact raises a question about generality of exponential behavior for problems of transport in random media. 
 %Standard large deviations approach, with fixed number of steps, dictates specific and model-based decay and hence non-applicable.
 By extending the Large Deviations approach for a continuous time random walk we uncover a general universal behavior for the decay of the density. 
 It is found that fluctuations in the number of steps of the random walker, performed at finite time, lead to exponential decay (with logarithmic corrections) of $P(X,t)$. 
 %Specifically $P(X,t)\sim\exp\left(-t \left[\kappa\log(|X|/t)^{1-1/\beta}|X|/t+C\right] \right)$ in the $|X|/t\to\infty$ limit ($\kappa$, $\beta$, and $C$ are constants). 
% Similar to the universal case of long times, dictated by the central limit theorem, when $P(X,t)\sim\exp\left(-t g(|X|/t) \right)$  and $g(y)\sim y^2$ ($y$ is small), we find that for any time the rate function $g(y)$ is almost linear (when $y$ is large). 
 This universal behavior 
 %is on the same footing as the central limit theorem, 
% (but for the tails)
 %and 
 holds also for short times, a fact that makes experimental observations %of such generalization of the central limit theorem for the tails 
 readily achievable.           
\end{abstract}

\maketitle

The emergence of normal Gaussian statistics for various random observables in nature is widespread. 
Examples range from chest sizes of Scottish soldiers~\cite{Scotts} to Brownian motion~\cite{Perrin}. 
This ``popularity" is attributed to the central limit theorem (CLT).   
The statement of the CLT is that, at its center, the distribution of  a sum of independent, identically distributed (IID), random variables is Gaussian. 
However, recently striking deviations from Gaussian behavior were recorded in a large number of experiments tracking spatial diffusion of tracer particles in various media. 
Interestingly, in many  measurements the observed probability density function (PDF), $P(X,t)$,  attains an exponential (or close to exponential) decay.
Examples include colloidal suspensions~\cite{WeitzGlass,Blaaderen}, nano particles in polymer solutions~\cite{Hue16}, molecular motion on a solid-liquid interface~\cite{Shwartz13,Shwartz17}, living cells~\cite{BioCell01},  phospholipid fluid tubules and biofilament networks~\cite{Granick01,Granick02}, active gels~\cite{MIzuno}, financial markets~\cite{Yakovenko}, colloidal glasses~\cite{Kob01,Weitz2010}, worms~\cite{Worms} and suspensions of swimming microorganisms~\cite{Goldstein01,Polin01} (more examples are provided in ~\cite{Granick01,ChechkinPRX}).

Appearance of a few jumps/excursions that dominate the process  is a common feature in a portion of these experiments. For a bead in a bacteria suspension such ``jumps" were attributed to temporal adhesion of the particle to a bacteria flow that establishes short-term motion alongside the bacteria~\cite{Polin01}. In the F-actin random network the displacements are myosin driven excursions~\cite{MIzuno}, while for Lennard-Jones suspensions there are ``cage-breaking" events~\cite{WeitzGlass}. 
The later system is one of four different (experimental and numerical) systems analyzed by Chaudhuri et al.~\cite{Kob01}. In this work the authors noticed that particle displacement measured in: dense suspension of colloidal hard spheres, slowly driven dense granular assembly, silica melt and a binary Lennard-Jones mixture, all show the same universal feature of exponential decay for $P(X,t)$ of the traced particles. Moreover, the authors used a special variant of the continuous time random walk  with two exponential PDFs for waiting times and two Gaussian distributions for the sizes of the jumps, in order to reproduce the observed behavior. 
These findings lead to questions regrading the universality of exponential decay. How can a basic random walk theory give rise to a theory that produces universal exponential decay for systems with  various distributions for jump sizes and waiting times? Is there a large class of processes that attains such universality? 
And if so, what is the precise mathematical description of the mentioned exponentional decay.
We wish to develop this theory and identify such broad class of processes by focusing on the  
 role of randomness of the number of jumps of a particle in an experiment.

In this manuscript we propose to reconcile observed non-Gaussianity by explicitly invoking the continuous time random walk (CTRW) formalism~\cite{MontrolWeis}, but without restricting ourselves to specific examples.
Specifically we extend the approach of large deviations to both space and time. The random number of measured jumps naturally occurs in CTRW due to random waiting times between the jumps. While for constant number of jumps the Large Deviations approach fails to produce universal behavior (as we show below), the situation with random number of jumps that we treat here is quite different.
%This formalism is a very widely applicable model of normal and anomalous transport~\cite{Bouchaud,KlafterMetzler,Masoliver}. 
We develop a subordination approach for Large Deviations and show that for any process that involves a random number of jumps, and can be modeled by CTRW, a very general statement holds: {\it the exponential tails for the positional PDF are rather a rule  and not an exception}.
The exponential decay of the tails is a general feature exactly like the Gaussian behavior (that is dictated by CLT) at the center.

Let us first stress out why, mathematically speaking, the numerously observed exponential decay is {\em unexpected} from the stand point of a regular random walk and standard Large Deviation approach~\cite{Touchete,Majumdar,Krapivsky,Dhar,TouchetteA,Derrida}. The random walk definition is as follows, at each step a particle can perform a step of size $x$, while the PDF of $x$ is given by $f(x)$. After $N$ steps the position $X$ is simply the sum of all random and independent steps $X=\sum_{i=1}^N x_i$. We will concentrate on the case when $f(x)$ is symmetric and decays as $f(x)\sim\exp\left(-(|x|/\delta)^\beta\right)$ when $|x|\to\infty$, ($\beta>1$). Namely, we exclude here power law decay of $f(x)$, in particular we are in the domain of attraction of the Gaussian CLT.
Indeed, the first moment  $\langle x\rangle$ is $0$ due to symmetry and the second moment, $\langle x^2\rangle=\int_{-\infty}^{\infty}x^2f(x)\,dx$, is finite. Thus according to CLT, when $|X|/N$ is not large, 
%and $N$ is large enough, 
the PDF to find the particle at position $X$ after $N$ steps, i.e. ${\cal P}_N(X)\sim\exp\left(-N[\frac{1}{\sqrt{2\langle x^2\rangle}} \left(\frac{|X|}{N}\right)]^2\right)$. All the different properties of $f(x)$ enter solely via  $\langle x^2 \rangle$, the functional Gaussian form is universal. The situation for the tails is quite different. According to the Theory of Large Deviations, and more specifically Cram{\'e}r's Theorem~\cite{Touchete}, for large $N\to\infty$, ${\cal P}_N(X)\sim\exp\left(-N I\left(\frac{|X|}{N}\right)\right)$ where the rate function 
$I(a)=\underset{\Omega}{\sup}\left\{\Omega a - \log\left[\langle \exp\left(\Omega x\right)\rangle\right]\right\}$. For small $|X|/N$, this leads to quadratic form of $I(|X|/N)$ and the already mentioned Gaussian behavior of ${\cal P}_N(X)$ at the center. 
But for the tails, a straight-forward result of this theorem is that when $|X|/N\to\infty$ the rate function takes the form $I(a)\sim a^\beta$ and
%%%%%%%%%%%%%%%%%%%%%%%%%%%%%%%%%%%%%%%%%%%%%%%%%%
\begin{equation}
    {\cal P}_N(X)\underset{N\to\infty}{\sim}
    \exp\left(-N\left[\frac{1}{\delta}\frac{|X|}{ N}\right]^\beta\right)
    \qquad\frac{|X|}{N}\to\infty.
    \label{largedevtail}
\end{equation}
%%%%%%%%%%%%%%%%%%%%%%%%%%%%%%%%%%%%%%%%%%%%%%%%%%%%%
%In \cite{SuppMat} we show that this result holds actually for any integer $N$ as long as $X$ is large enough. 
Indeed, if $\beta=2$, ${\cal P}_N(X)$ is Gaussian, and hence exponential tails are not present.
From Eq.~(\ref{largedevtail}) it becomes quite obvious that the functional form of the decay is $\beta$ dependent and non-universal. It is natural to expect that the decay of the tails is very specific. This is why, based on the regular random walk perspective, the large number of different experiments that show very similar functional decay (i.e. exponential) is definitely unexpected. Unless one assumes intrinsic exponential distribution for the jumps of the particle ($\beta=1$), in all the different experiments, which is unlikely.

We already mentioned our intention to resolve this issue by using the fact that the number of steps in most experiments is random for any finite measurement time $t$. Probably the simplest assumption is that the particle will wait for some random time $\tau$ between successive steps. 
This assumption is exactly the framework of CTRW and it leads to randomization of  $N$~\cite{Bouchaud}. 
The CTRW is a widely applicable model for transport in disordered media~\cite{Bouchaud,KlafterMetzler} that  describes a particle that performs random independent steps $x$, determined by the PDF $f(x)$, and between two successive steps the particle waits a random time $\tau$ that is distributed according to $\psi(\tau)$. All the waiting times are independent.   
%Further we assume that all the different waiting times between the steps are independent and identically distributed according to $\psi(\tau)$. 
The probability of observing $N$ steps at time $t$, $Q_t(N)$, is fully determined by $\psi(\tau)$ (see below).
%and the independence assumption. 
For CTRW the position $X=\sum_{i=1}^N x_i$ depends both on the random  $\{x_i\}$s and the random $N$.
%, i.e. a random sum of random variables. 
By conditioning on the specific outcomes of $N$ steps, the PDF to find the particle at $X$ at time $t$ is 
%%%%%%%%%%%%%%%%%%%%%%%%%%%%%%%%%%%%%%%%%%%%%%
\begin{equation}
    P(X,t)=\sum_{N=0}^\infty {\cal{P}}_N(X) Q_t(N). 
\label{subordinGen}
\end{equation}
%%%%%%%%%%%%%%%%%%%%%%%%%%%%%%%%%%%%%%%%%%%%%%
Eq.~(\ref{subordinGen}) is also known as the subordination of the spatial process for $X$ by the temporal process for $N$~\cite{Bouchaud,BurovBarkai,Burov2017,MarcinSub,Sokolov08}. The regular approach for CTRW without anomalously large jumps~\cite{KlafterMetzler} is to replace ${\cal P}_N(X)$ in Eq.~(\ref{subordinGen}) by the Gaussian approximation.
From Eq.~(\ref{largedevtail}) it is clear that for the tails, the Gaussian approximation is simply incorrect, unless $\beta=2$. In order to accomplish the calculation of $P(X,t)$ for large $|X|$ the first thing to do is to insert the form of ${\cal P}_N(X)$ in Eq.~(\ref{largedevtail}) into Eq.~(\ref{subordinGen}), instead of the Gaussian approximation, i.e. subordination of Large Deviations. 
In the following we provide a sketch of the proof while a complete derivation will be published in a longer publication.
In Eq.~(\ref{subordinGen}) one notices that for large $|X|$ the form of Large Deviations for ${\cal P}_N(X)$ (i.e. Eq.~(\ref{largedevtail})) states that all the small $N$ contributions of ${\cal P}_N(X)$ are negligible, as compared to large $N$. %(Eq.~(\ref{largedevtail}) holds for any $N$).
So the sum in Eq.~(\ref{subordinGen}), with Eq.~(\ref{largedevtail}) for ${\cal P}_N(X)$, is affected only by large $N$ values of $Q_t(N)$ when $|X|$ is large.
Indeed, for any fixed $t$ the position $|X|$ can be chosen arbitrary large in-order to suppress all the contributions of ${\cal{P}}_N(X) Q_t(N)$, for any finite $N$. 
It is thus crucial to obtain the large $N$ behavior of $Q_t(N)$.

%%%%%%%%%%%%%%%%%%%%%%%%%%%%%%%

\begin{figure} 
%\centering
\begin{center}
			\includegraphics[width=0.42\textwidth]{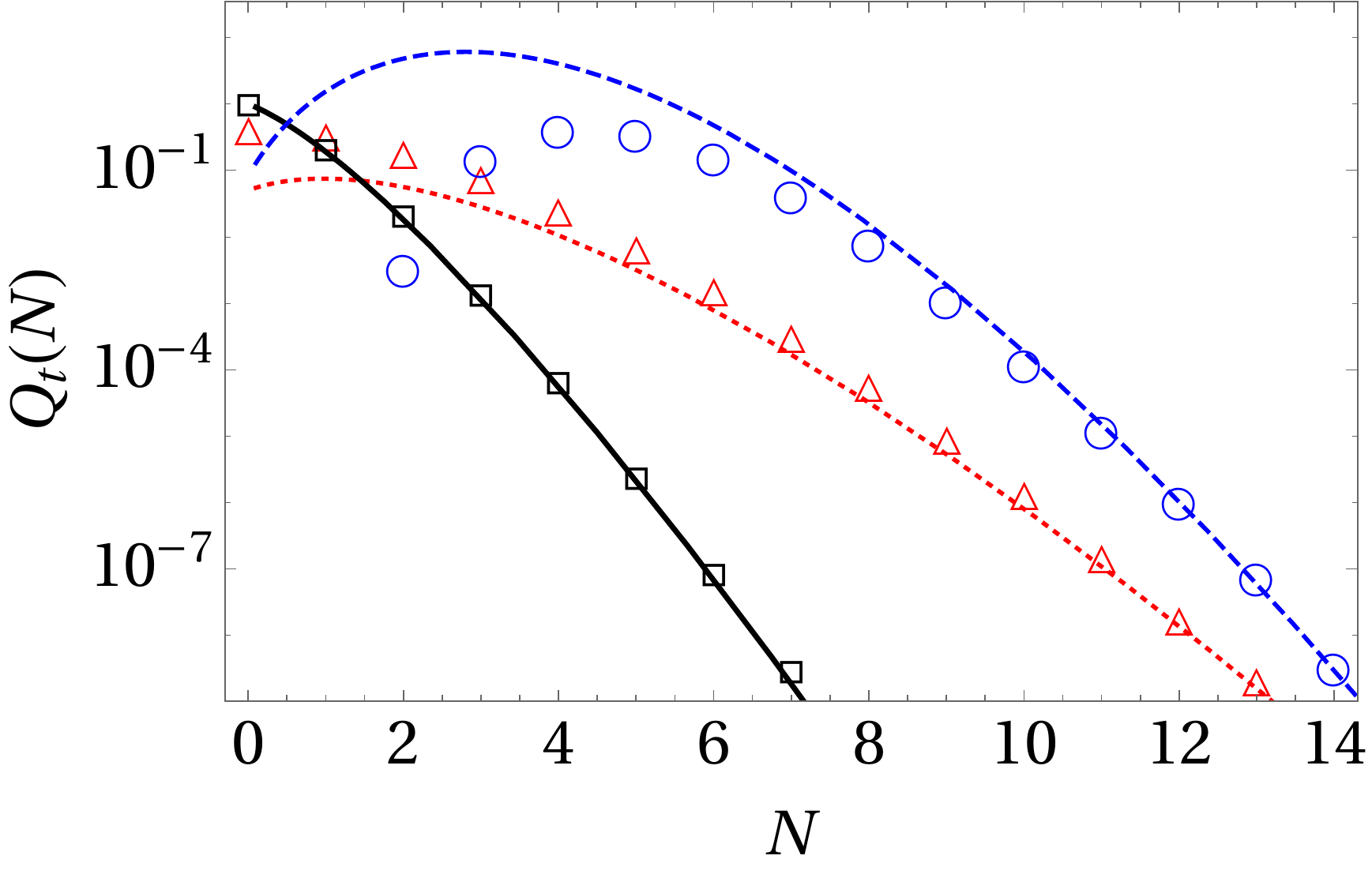}
\end{center}
\caption{
Numerical simulations (symbols) of $Q_t(N)$ are compared to Eq.~(\ref{qtnlimit}) (lines) for three different $\psi(\tau)$s. {\color{black}$\Box$} is the Half Gaussian Distribution $\psi(\tau)=\frac{2}{5\pi}\exp(-\tau^2\big/25\pi)$, measurement time is $t=1.5$. {\color{blue}$\bigcirc$}  is a special form of Beta distribution $\psi(\tau)=6\tau(1-\tau)$, $0\leq\tau\leq1$ ($t=1.5$), and {\color{red}$\bigtriangleup$} is the Dagum distribution $\psi(\tau)=1/(1+\tau)^2$ while measurement time $t=2.5$.
}
    \label{qtnexample}
\end{figure}
%%%%%%%%%%%%%%%%%%%%%%%%%%%%%%%%%

{\it {$Q_t(N)$ in the $N\to\infty$ limit}} is the probability of occurrence of a rare event, i.e., large number of steps, in finite time. 
The distribution of the dwell time $\tau$ between two steps, $\psi(\tau)$, is independent of previous or following waiting times. 
%Each time a step is performed, the waiting process is renewed ( since the name a renewal process~\cite{CoxBook}). 
The probability $Q_t(N)$ is the probability that 
$\sum_{i=1}^N\tau_i<t$ while $\sum_{i=1}^{N+1}\tau_i>t$, where $\{\tau_i\}$ are the waiting times.
%a sum of $N$ positive independent random variables $\tau_i$ equals to $t'<t$, and the $N+1$ random variable $\tau_{N+1}$ is larger than $t-t'$. 
Due to convolution property of Laplace transform, ${\hat{Q}}_s(N)=\int_0^\infty\exp\left(-st\right)Q_t(N)\,dt$ is~\cite{Godreche,CoxBook}
%%%%%%%%%%%%%%%%%%%%%%%%%%%%%%%%%%%%%%%%%%%%%%%%%
\begin{equation}
 {\hat{Q}}_s(N)= {\hat{\psi}}(s)^N\left(1-{\hat{\psi}}(s)\right)\big/s  ,
  \label{renewalEq}
\end{equation}
where ${\hat{\psi}}(s)=\int_0^\infty\psi(t)\exp\left(-st\right)\,dt$.
It is assumed that the short time ($\tau\to 0$) Taylor expansion of $\psi(\tau)$ is
%%%%%%%%%%%%%%%%%%%%%%%%%%%%%%%%%%%%%%%%%%%%%%%%%
\begin{equation}
    \psi(\tau)\underset{\tau\to 0}{\sim} \sum_{j=0}^\infty C_{A+j}\tau^{A+j},
    \label{psiexpansion}
\end{equation}
%%%%%%%%%%%%%%%%%%%%%%%%%%%%%%%%%%%%%%%%%%%%%%%%%%
where $A\geq 0$ is an integer. 
This is a very natural assumption as it merely demands that $\psi(\tau)$ will be analytic at the vicinity of $\tau=0$.
In the Appendix~\ref{appendixA} we use a power series expansion of ${\hat{Q}}_s(N)$ to show that the leading term of $Q_t(N)$ in the large $N$ limit is 
%%%%%%%%%%%%%%%%%%%%%%%%%%%%%%%%%%%%%%%%%%%%%%%%
\begin{equation}
      Q_t(N)\underset{N\to\infty}{\sim}
    \frac{\left(\left[C_A\Gamma(A+1)\right]^\frac{1}{A+1} t \right)^{N(A+1)}}{\Gamma\left[N(A+1)+1\right]}
e^{\frac{C_{A+1}}{C_A}t}.
%    \begin{cases}
 %   e^{\frac{C_{A+1}}{C_A}t}&  m= 1\\
  %  1              & m>1
%\end{cases}
    \label{qtnlimit}
\end{equation}
%%%%%%%%%%%%%%%%%%%%%%%%%%%%%%%%%%%%%%%%%%%%%%%%%%
It attains the form of the large deviation principle, $Q_t(N)\sim\exp\left(-N I_T(t/N)\right)$ (see Appendix~\ref{appendixA}).
This general result holds for any $\psi(\tau)$, 
while $t$ is kept constant and $N\to\infty$. It is not affected by the large $\tau$ behavior of $\psi(\tau)$ and includes situations when $\langle\tau\rangle\to\infty$, i.e. anomalous diffusion~\cite{Bouchaud,KlafterMetzler} .
 For the case when $\psi(\tau)$ is exponential, $Q_t(N)$ is a Poisson distribution and Eq.~(\ref{qtnlimit}) agrees perfectly with this fact.

%%%%%%%%%%%%%%%%%%%%%%%%%%%%%%%%%%%%%%%%%%%%%%%%%%%%%%%%%%%%%%%%%%%%%%%%%%%%%%
%%%%%%%%%%%%%%%%%%%%%%%%%%%%%%%%%%%%%%%%%%%%%%%%%%%%%%%%%%%%%%%%%%%%%%%%%%
%%%%%%%%%%%%%%%%%%%%%%%%%%%%%%%%%%%%%%%%%%%%%%%%%%%%%%%%%%%%%%%%%%%%%%%%%%

\begin{figure} 
\centering
			\includegraphics[width=0.4\textwidth]{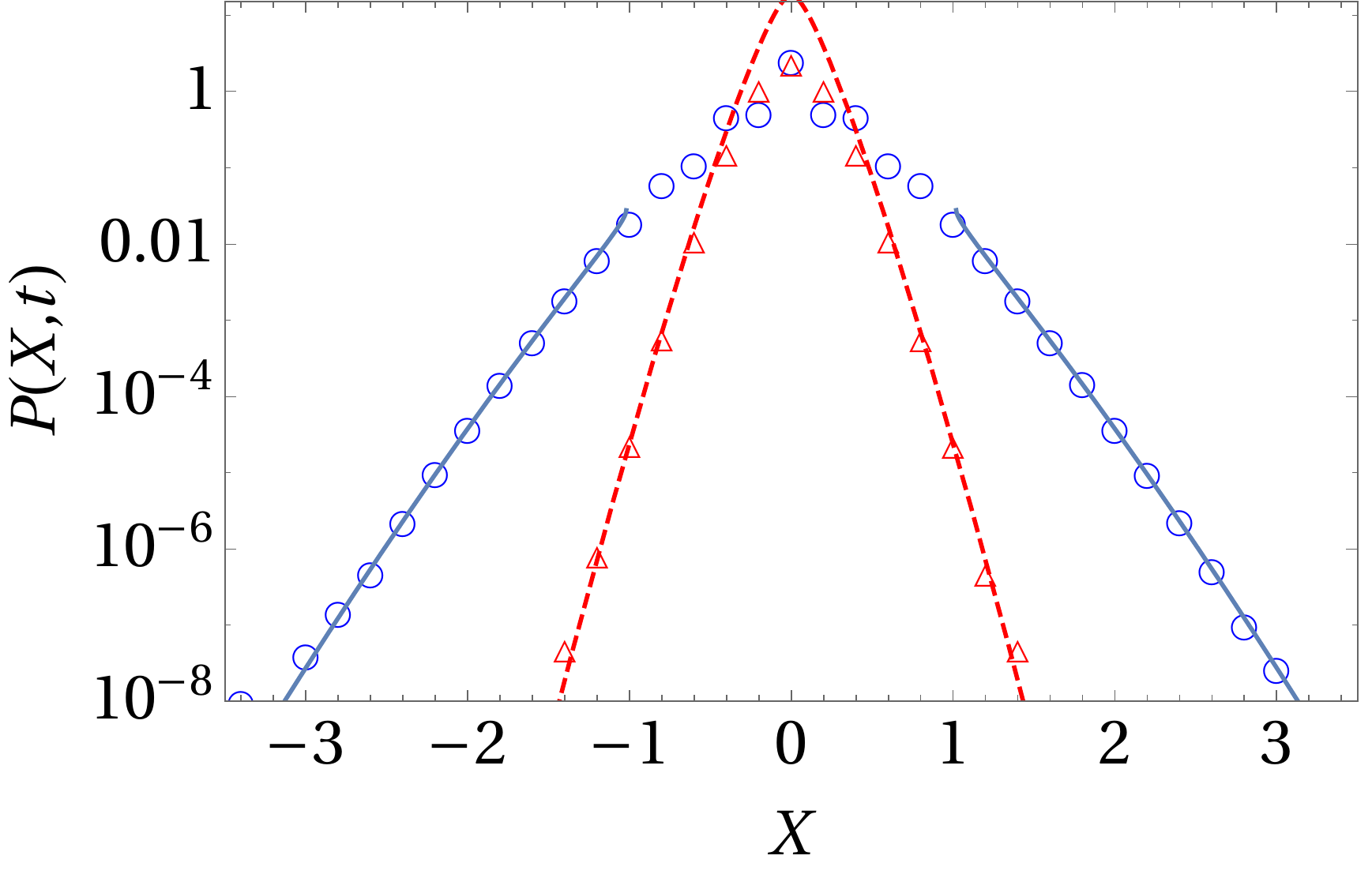}
\caption{
Universality of exponential tails. Comparison of $P(X,t)$ obtained from CTRW simulations (symbols) and theoretical prediction Eq.~(\ref{pxtsolution}) without fitting (see Appendix~\ref{appendixB}). {\color{blue}$\bigcirc$} displays CTRW where $f(x)$ is uniform between $-0.5$ and $0.5$ (and zero everywhere else) , $\psi(\tau)=1\big/(1+\tau)^2$, $t=1.5$, and the theory is the thick line. {\color{red}$\bigtriangleup$} is for $f(x)=\exp(-50 x^2)\big/\sqrt{\pi/50}$, $\psi(\tau)$ is uniform between $0$ and $1$ (and zero everywhere else), $t=1.5$, and the theory is the dashed line. Notice the log-linear coordinates, indicating that $P(X,t)$ is exponential like in many experiments mentioned in the introduction. 
}
    \label{pxtexample}
\end{figure}

Supplemented with the general result for $Q_t(N)$ and using Eq.~(\ref{largedevtail}) for ${\cal P}_N(X)$ we finally obtain the tail behavior of $P(X,t)$. We plug Eq.~(\ref{largedevtail}) and Eq.~(\ref{qtnlimit}) into Eq.~(\ref{subordinGen}),  approximate the sum by an integral over $N$ and obtain 
\vspace{0.5cm}
%%%%%%%%%%%%%%%%%%%%%%%%%%%%%%%%%%%%%%%%%%%%%%%%%%%%%%%
\begin{widetext}
\begin{equation}
   % \begin{array}{ll}
       P(X,t)\sim     
       %\\
     \int_0^\infty
    %\frac{ 
    \exp{\left(
    \overbrace{-N(
    \left[\frac{1}{\delta}\frac{|X|}{ N}\right]^\beta-\frac{C_{A+1}}{C_A}\frac{t}{N}
    -(A+1)\left[\log(\frac{(C_A\Gamma(A+1))^{\frac{1}{A+1}}}{A+1}\frac{t}{N})+1\right]
    )}^{K(N)}
  \right)}
  %  }{\Gamma\left(N(A+1)+1\right)}
    \,dN    .
%    \end{array}
        \label{integralPxt}
\end{equation}
\end{widetext}
%%%%%%%%%%%%%%%%%%%%%%%%%%%%%%%%%%%%%%%%%%%%%%%%%%%%%%%%
Clearly this represents a subordiantion of Large Deviations result, i.e. Cramer's theorem with the just obtained  universal $Q_t(N)$.
We now use the saddle point approximation in order to calculate the integral for $|X|\to\infty$. 
We find that the maximum of $K(N)$
is achieved for 
%%%%%%%%%%%%%%%%%%%%%%%%%%%%%%%%%%%%%
\begin{equation}
    N^*=|X| 
    g_0 W_0\left[g_1\left( \frac{|X|}{t}\right)^\beta\right]^{-1/{\beta}}
\end{equation}
%%%%%%%%%%%%%%%%%%%%%%%%%%%%%%%%%%%%%%%%%
where $g_0=(\beta(\beta-1)\big/(A+1))^{1/\beta}/\delta$ , $g_1=\left[g_0(A+1)\big/(C_A\Gamma(A+1))^{\frac{1}{A+1}}\right]^\beta$ and $W_0(y)$ is the principal branch of a Lambert $W$ function~\cite{lambertMath,Corless,Meerson,Lior2018}, i.e. a solution of the equation $W(y)\exp\left(W(y)\right)=y$.
Therefore the asymptotic behavior of $P(X,t)$ in the $|X|\to\infty$ limit is provided by 
%%%%%%%%%%%%%%%%%%%%%%%%%%%%%%%%%%%%%%%%%%%%%%
\begin{equation}
    P(X,t)\underset{|X|\to\infty}{\sim} 
    \frac{\exp\left(
    -t\left\{
    \frac{|X|}{t}Z\left(\frac{|X|}{t}\right)+C
    \right\}
    \right)}{\sqrt{2\pi K''(N^*)}}
            \label{pxtsolution}
\end{equation}
%%%%%%%%%%%%%%%%%%%%%%%%%%%%%%%%%%%%%%%%%%%%%%%
where 
%%%%%%%%%%%%%%%%%%%%%%%%%%%%%%%%%%%%%%%%%%%%%%%%%
\begin{equation}
 Z(y)=\frac{\left(\frac{g_0(A+1)}{\beta}+\frac{1}{g_0^{\beta-1}\delta^\beta}\right)W_0\left[g_1y^\beta\right]-g_0(A+1)}{W_0\left[g_1y^\beta\right]^{\frac{1}{\beta}}}
    \label{ztdefinition}
\end{equation}
%%%%%%%%%%%%%%%%%%%%%%%%%%%%%%%%%%%%%%%%%%%%%%%%%
and $C=-C_{A+1}\big/C_A$.
The function $W_0(y)$ ($y\geq-1/e$) is a monotonically increasing function with sub-logarithmic slow growth, 
$\log(y)-\log(\log(y))\leq W_0(y)\leq \log(y)-\frac{1}{2}\log(\log(y))$ for $e\leq y$~\cite{Lambertineq}.
The asymptotic expansion for $y\to\infty$ is $W_0(y)\sim\log(y)-\log\left(\log(y)\right)$ and in the limit $|X|/t\to\infty$, Eq.~(\ref{pxtsolution}) obtains the form 
%%%%%%%%%%%%%%%%%%%%%%%%%%%%%%%%%%%%%%%%%%%%%%%%%
\begin{equation}
    P(X,t)\underset{\frac{|X|}{t}\to\infty}{\sim}
     \exp \left(-t\left[\kappa\log\left(\frac{|X|}{t}\right)^{1-1/\beta}\frac{|X|}{t}+C\right]  \right)
\label{asymptpxt}
\end{equation}
%%%%%%%%%%%%%%%%%%%%%%%%%%%%%%%%%%%%%%%%%%%%%%%%%
and $\kappa=\left(g_0(A+1)/\beta+1/{g_0^{\beta-1}\delta^\beta}\right)\beta^{1-1/\beta}$. This result states that the tails of $P(X,t)$ will exhibit almost exponential decay. The logarithmic corrections, due to slow sub-logarithmic growth of $Z(\dots)$ in Eq.~(\ref{pxtsolution}),
will cause small deviations from pure exponential behavior and overall it would seem that $yZ(y)+C$ converges to linear form. 
In Fig~\ref{pxtexample} this (approximately) exponential behavior  of $P(X,t)$ is displayed for two different pairs of $f(x)$ and $\psi(\tau)$. 
As we already mentioned, in the case of $P(X,t)$, $t$ is not limited to the domain of large values. 
If $t$ can take small enough values, while keeping the values of $Z(|X|/t) |X|/t+C$ not too large, the exponential behavior can be readily observed in experimental situation.

%%%%%%%%%%%%%%%%%%%%%%%%%%%%%%%

%%%%%%%%%%%%%%%%%%%%%%%%%%%%%%%%%

\begin{figure} 
\centering
			\includegraphics[width=0.4\textwidth]{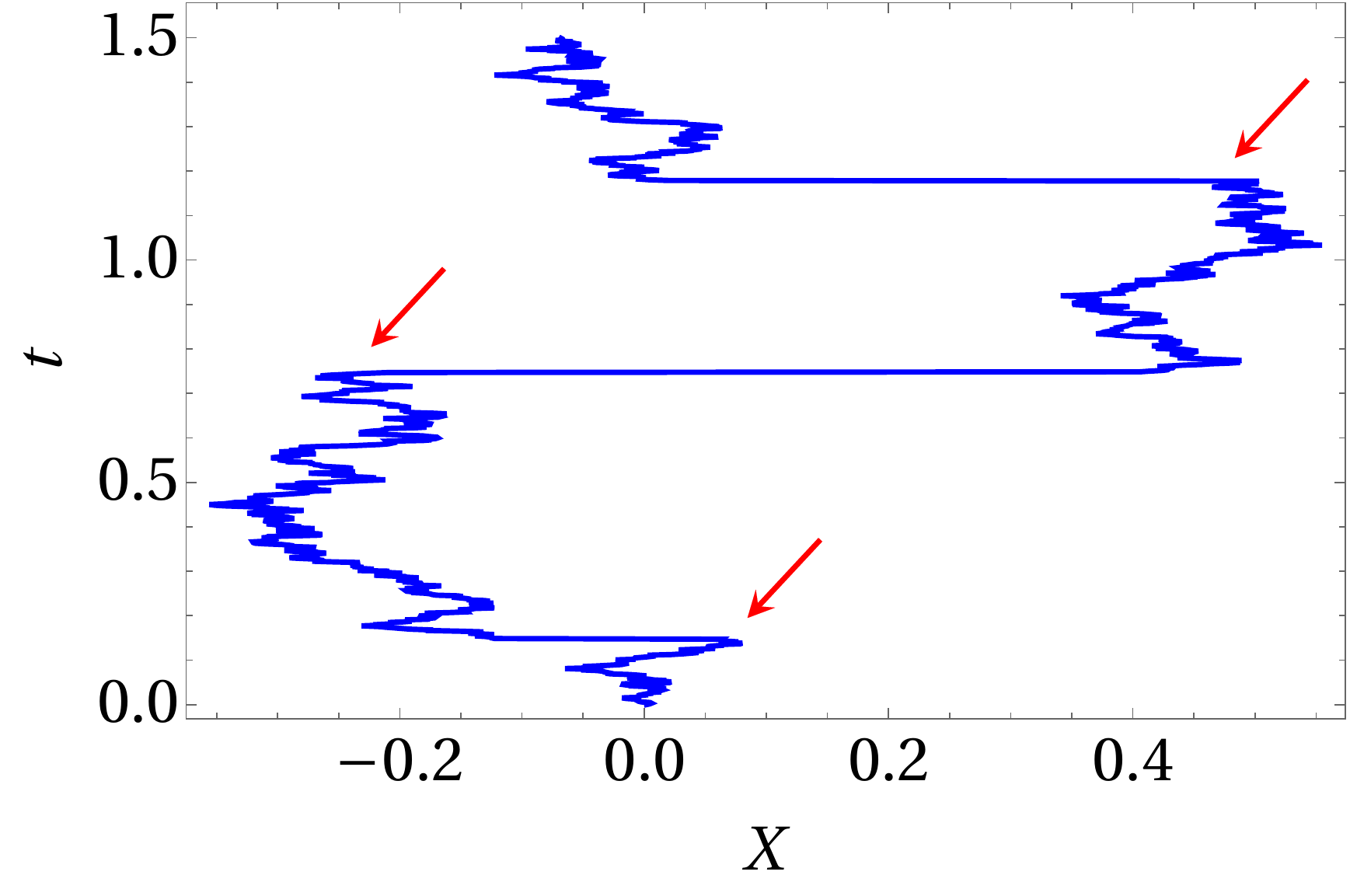}
\caption{
Typical trajectory of Brownian motion combined with CTRW. The arrows label the instantaneous jumps of the CTRW process. The probability to observe the particle at large $X$ is controlled by appearence of such jumps.
}
    \label{trajexample}
\end{figure}

This explains the large amount of experimental systems where the exponential decay of the $P(X,t)$ was recorded. Specifically relevant are the glassy systems~\cite{Ciamarra}, where the CTRW approach ~\cite{Kob01,Blumen,Ciamarra}, proved to be useful. The dynamics of single glass-formers, e.g. colloids, often presents itself in the form of ``jumps" and those jumps are also present in many of the glass theories, as discussed in~\cite{Ciamarra}.
%such as the mode-coupling, free volume and the random first order theory~\cite{Ciamarra}. 
Such jumps will usually look like extreme events on top of a caged Brownian motion, e.g. Fig.~\ref{trajexample}. As long as the scale of those very fast and rare ``jumps" is significantly larger than the ongoing diffusion, the tails of $P(X,t)$ will be completely dictated by the statistics of those ``jumps".   
The randomness in the number of such $N$ ``jumps" is what makes it an important factor behind the observed universal exponential decay of $P(X,t)$ and the convergence to exponential behavior occurs even on time scales when the average number of jumps is small.

When the exponential decay of the tails of $P(X,t)$ is compared to Gaussian behavior at the center, it is important to stress out the different time-scales when these two behavior will take place. The Gaussian behavior will appear only when the measurement time is sufficiently long, while the exponential decay will take place for any time (as we already mentioned). Both features are based on statistics of large number of events. While for the center of $P(X,t)$, large number of events is sampled only for long enough time, the tails that describe the rare events are by themselves a manifestation of appearance of a large group of events. It is then expected that in an experimental situation the exponential decay will show itself long before the convergence to Gaussian will appear~\cite{Granick02}.
For long measurement time Eq.~(\ref{pxtsolution}) also holds, but when $t$ is large   $|X|$ must be enormous in-order for $|X|/t$ to be sufficiently large. Thus the exponential tails are simply pushed towards really small values of $P(X,t)$, i.e. far from the center.

%%%%%%%%%%%%%%%%%%%%%%%%%%%%%%%

%%%%%%%%%%%%%%%%%%%%%%%%%%%%%%%%%

We presented a space-time theory for large deviations of the widely applicable continuous time random walk. This theory provides an explanation for a large class of recent experimental observations of diffusion processes. 
In this sense, the reported universal behavior is likely to establish a link between experiments and the theory of large deviations. 
The large deviation principle for space, ${\cal P}_N(X)\sim\exp[-N I_S(\frac{X}{N})]$ Eq.~(\ref{largedevtail}), and for time $Q_t(N)\sim\exp[-N I_T(\frac{t}{N})]$ Eq.~(\ref{qtnlimit}), were described by rate functions $I_S(\dots)$ and $I_T(\dots)$, respectively. 
The subordination approach yielded our main result, Eq.~(\ref{pxtsolution}), where the rate function $\frac{|X|}{t} Z\left(\frac{|X|}{t}\right)+C$ controls the decay of the PDF. 
It is however remarkable that our theory works for any $t$. This stems from the fact that once $X$ is large, a large number of jumps is needed to arrive to this position. When the number of jumps is fixed, as in a standard random walk, the widely observed universal decay is completely missed. Deviations from the presented theory are expected when $\psi(\tau)$ is non-analytic in the vicinity of $\tau=0$ or when the decay of $f(x)$ is broad, e.g. power-law.

It is also worth mentioning that the CTRW formalism will present a Fickian diffusion, i.e. linear growth of the mean squared distance with time, as long as $\langle\tau\rangle$ is finite~\cite{Kehr}.
%and the initial positions are sampled from an ensemble of particles. 
This means that the presented broad class of models investigated here will also show the widely investigated  Fickian yet non-Gaussian behavior~\cite{Granick01,Granick02,Chubynsky,Sebastian,ChechkinPRX,Akimoto}. 
Finally, we notice that the presented results are expected to have a high impact on the field of triggered reactions in Physics, Chemistry and Biology~\cite{Loverdo,Benichou,Grebenkov}. In any situation where  a reaction occurs as a result of a first arrival, the universal rare behavior described here will dominate due to the simple fact that exponential decay is significantly slower than the Gaussian case of simple diffusion.  This has crucial consequences for transport in such systems as the living cell~\cite{Grebenkov,Tabei}.

{\bf Acknowledgments:} This work was supported by the Pazy foundation grant 61139927. EB acknowledges the Israel Science Foundation’s grant
1898/17.

\section{Appendix}

\subsection{Derivation of $Q_t(N)$}
\label{appendixA}
In the  article we presented the problem of observing $N$ jumps for a renewal process when the waiting times between the jumps are random and distributed according to $\psi(\tau)$. We use the assumption that the process posses the renewal property, i.e. independent $\tau$ and that the Taylor expansion of $\psi(\tau)$ in the $\tau\to 0$ limit is provided by
%%%%%%%%%%%%%%%%%%%%%%%%%%%%%%%%%%%%%%%%%%%%%%%%%%%
\begin{equation}
\psi(\tau)\underset{\tau\to 0}{\sim}\sum_{i=0}^\infty C_{A+i}\tau^{A+i}.
    \label{taylorexpansion}
\end{equation}
%%%%%%%%%%%%%%%%%%%%%%%%%%%%%%%%%%%%%%%%%%%%%%%%%%%%
The probability $Q_t(N)$ is the probability that a sum of $N$ positive independent random variables $\tau_i$ equals to $t'<t$, and the $N+1$ random variable $\tau_{N+1}$ is larger than $t-t'$. Due to convolution property of Laplace transform, ${\hat{Q}}_s(N)=\int_0^\infty\exp\left(-st\right)Q_t(N)\,dt$ is~\cite{Godreche,CoxBook}
%%%%%%%%%%%%%%%%%%%%%%%%%%%%%%%%%%%%%%%%%%%%%%%%%
\begin{equation}
 {\hat{Q}}_s(N)= {\hat{\psi}}(s)^N\left(1-{\hat{\psi}}(s)\right)\big/s  ,
  \label{renewalEq}
\end{equation}
where ${\hat{\psi}}(s)=\int_0^\infty\psi(\tau)\exp\left(-s\tau\right)\,d\tau$. Eq.~(\ref{taylorexpansion}), and Tauberian Theorem~\cite{Weiss}, dictates the form of ${\hat{\psi}}(s)$ in the limit $s\to\infty$
%%%%%%%%%%%%%%%%%%%%%%%%%%%%%%%%%%%%%%%%%%%%
\begin{equation}
    {\hat{\psi}}(s)\underset{s\to\infty}{\sim}\sum_{i=0}^\infty\frac{C_{A+i}\Gamma\left(A+i+1\right)}{s^{A+i+1}}.
    \label{tauberianpsi}
\end{equation}
%%%%%%%%%%%%%%%%%%%%%%%%%%%%%%%%%%%%%%%%%%%%
By introducing Eq.~(\ref{tauberianpsi}) into Eq.~(\ref{renewalEq}) we obtain the $s\to\infty$ expansion of ${\hat{Q}}_s(N)$
%%%%%%%%%%%%%%%%%%%%%%%%%%%%%%%%%%%%%%%%%%%%%%%%
\begin{equation}
{\hat{Q}}_s(N)\sim\left(\frac{C_A\Gamma(A+1)}{s^{A+1}}+
\sum_{i=1}^\infty\frac{C_{A+i}\Gamma(A+i+1)}{s^{A+i+1}}\right)^N
\left(\frac{1}{s}-\sum_{i=0}^\infty\frac{C_{A+i}\Gamma(A+i+1)}{s^{A+i+2}}\right),
    \label{qsnexp01}
\end{equation}
%%%%%%%%%%%%%%%%%%%%%%%%%%%%%%%%%%%%%%%%%%%%%%%
that is easily rewritten as
%%%%%%%%%%%%%%%%%%%%%%%%%%%%%%%%%%%%%%%%%%%%%%%%
\begin{equation}
{\hat{Q}}_s(N)\sim\frac{\left[C_A\Gamma(A+1)\right]^N}{s^{N(A+1)+1}}\left(1+
\sum_{i=1}^\infty\frac{C_{A+i}\Gamma(A+i+1)}{C_A \Gamma(A+1) s^{i}}\right)^N
\left(1-\sum_{i=0}^\infty\frac{C_{A+i}\Gamma(A+i+1)}{s^{A+i+1}}\right),
    \label{qsnexp02}
\end{equation}
%%%%%%%%%%%%%%%%%%%%%%%%%%%%%%%%%%%%%%%%%%%%%%%
and by application of the binomial expansion 
%%%%%%%%%%%%%%%%%%%%%%%%%%%%%%%%%%%%%%%%%%%%%%%%
\begin{equation}
{\hat{Q}}_s(N)=\frac{\left[C_A\Gamma(A+1)\right]^N}{s^{N(A+1)+1}}\sum_{n=0}^N\binom{N}{n}\left(\sum_{i=1}^\infty\frac{C_{A+i}\Gamma(A+i+1)}{C_A \Gamma(A+1) s^{i}}\right)^n
\left(1-\sum_{i=0}^\infty\frac{C_{A+i}\Gamma(A+i+1)}{s^{A+i+1}}\right).
    \label{qsnexp03}
\end{equation}
%%%%%%%%%%%%%%%%%%%%%%%%%%%%%%%%%%%%%%%%%%%%%%%
Next we use the multinomial expansion 
$(\sum_{i=0}^m w_i )^n= \sum_{\{k_i\}}\frac{n!}{{ \prod_{i=0}^m k_i!}}\prod_{i=0}^m w_i^{k_i}$ where the summation is for all possible variations of integer $k_i$ such that the condition $\displaystyle\sum_{i=0}^m k_i=n$ holds. We use this form in the $m\to\infty$ limit and obtain from Eq.~(\ref{qsnexp03})
%%%%%%%%%%%%%%%%%%%%%%%%%%%%%%%%%%%%%%%%%%%%%%%%
\begin{equation}
\begin{array}{l}
{\hat{Q}}_s(N)=\displaystyle \sum_{n=0}^N\sum_{\{ k_i\}}\sum_{j=0}^\infty
\binom{N}{n} 
\frac{\left[C_A\Gamma(A+1)\right]^N}{s^{N(A+1)+1}}\times
\\
\times \displaystyle
\frac{n!}{{\displaystyle \prod_{i=0}^\infty k_i!}}\left(\prod_{i=0}^\infty \left[\frac{C_{A+1+i}\Gamma(A+2+i)}{C_A\Gamma(A+1)}\right]^{k_i}\right)\frac{B_j}{s^{\sum_{i=0}^\infty(i+1)k_i+l_j}},
\end{array}
    \label{qsnexp04}
\end{equation}
%%%%%%%%%%%%%%%%%%%%%%%%%%%%%%%%%%%%%%%%%%%%%%%
where $B_0=1$ and $l_0=0$ while for any $j\geq 1$ $B_j=-C_{A+j-1}\Gamma(A+j)$ and $l_j=A+j$. The inverse Laplace transform of $1/s^{\alpha+1}$ is $t^\alpha\big/\Gamma(\alpha+1)$, when $\alpha>-1$, then from Eq.~(\ref{qsnexp04}) the form of $Q_t(N)$ is 
%%%%%%%%%%%%%%%%%%%%%%%%%%%%%%%%%%%%%%%%%%%%%%%%
\begin{equation}
\begin{array}{l}
{{Q}}_t(N)=
\displaystyle
\sum_{n=0}^N\sum_{\{ k_i\}}\sum_{j=0}^\infty
\frac{\left[C_A\Gamma(A+1)\right]^N N!}{(N-n)!{\displaystyle \prod_{i=0}^\infty k_i!}}
\left(\prod_{i=0}^\infty \left[\frac{C_{A+1+i}\Gamma(A+2+i)}{C_A\Gamma(A+1)}\right]^{k_i}\right)\times
\\ \displaystyle
\times
\frac{B_j}{\displaystyle \Gamma(N(A+1)+\sum_{i=0}^\infty(i+1)k_i+l_j+1)}
t^{ \displaystyle \{  N(A+1)+\sum_{i=0}^\infty(i+1)k_i+l_j \} }.
\end{array}
    \label{qtnlong}
\end{equation}
%%%%%%%%%%%%%%%%%%%%%%%%%%%%%%%%%%%%%%%%%%%%%%%%
By taking out of the sum the common multiplier $\frac{\left(C_A\Gamma(A+1)t^{A+1}\right)^N}{\Gamma(N(A+1)+1}$, Eq.~(\ref{qtnlong}) is transformed into
%%%%%%%%%%%%%%%%%%%%%%%%%%%%%%%%%%%%%%%%%%%%%%%%
\begin{equation}
\begin{array}{l}
{{Q}}_t(n)=
\displaystyle
\frac{\left(C_A\Gamma(A+1)t^{A+1}\right)^N}{\Gamma(N(A+1)+1)}
\sum_{n=0}^N\sum_{\{ k_i\}}\sum_{j=0}^\infty
B_j
t^{ \displaystyle \left[  \sum_{i=0}^\infty(i+1)k_i+l_j \right] }\times
\\
\times
\displaystyle
\left(\prod_{i=0}^\infty \left[\frac{C_{A+1+i}\Gamma(A+2+i)}{C_A\Gamma(A+1)}\right]^{k_i}\right)
\frac{\left(N(A+1)\right)! N!n!}{n!(N-n)!{\displaystyle \prod_{i=0}^\infty k_i!}}\times
\frac{1}{\displaystyle \Gamma\left(N(A+1)+\sum_{i=0}^\infty(i+1)k_i+l_j+1\right)}.
\end{array}
    \label{qtnlong02}
\end{equation}
%%%%%%%%%%%%%%%%%%%%%%%%%%%%%%%%%%%%%%%%%%%%%%%%

We now fix $t$ and take the $N\to\infty$ limit. 
The prefactor $\frac{\left(C_A\Gamma(A+1)t^{A+1}\right)^N}{\Gamma(N(A+1)+1}$ in Eq.~(\ref{qtnlong02}) is the leading order term in $N$ and the only other terms in the triple sum that contribute to the leading order are the ones that do not converge to $0$ as $N\to\infty$. The next step is to identify the terms with the largest contribution to the value of the sum in the $N\to\infty$ limit.

First of all, since $\forall j$, $l_j\geq 0$ and $B_j$ is independent of $N$, we should take the minimal value of $l_j$, i.e. $j=0$. For any $j>0$, $l_j>0$, and such terms differ from the $j=0$ case by multiplications by terms of the form $\displaystyle 1\big/\prod_{r=1}^{A+j}\left(N(A+1)+\sum_{i=0}^\infty(i+1)k_i+r\right)$. Consequently, the summation over $j$ in Eq.~(\ref{qtnlong02}) contributes only the $j=0$ term. 

Next we treat the summation over all different $k_i$, i.e $\sum_{\{k_i\}}$. 
For a given $n$ and a given realization of $\{k_i\}$ the term that is needed to be considered is
%%%%%%%%%%%%%%%%%%%%%%%%%%%%%%%%%%%%%%%%%%%%%%%%%%%
\begin{equation}
\displaystyle \frac{\prod_{i=0}^\infty \left[\frac{C_{A+1+i}\Gamma(A+2+i)}{C_A\Gamma(A+1)}\right]^{k_i}}
{\displaystyle \left(\prod_{i=0}^\infty k_i!\right)\Gamma\left( N(A+1)+\sum_{i=0}^\infty(i+1)k_i+1\right)}.
    \label{termsumi}
\end{equation}
%%%%%%%%%%%%%%%%%%%%%%%%%%%%%%%%%%%%%%%%%%%%%%%%%%%%
Due to the fact that $\sum_{i=0}^\infty k_i=n$ it is obvious that $1\Big/\Gamma(\dots)$ is maximal for the realization when $k_i=n\delta_{i,0}$, ($\delta_{i,j}$ is Kronicker $\delta$-function). Any other realization of $\{k_i\}$ will introduce multiplication by $\sum_{i=0}^\infty i k_i\geq 1$ terms of the form 
$1\big/\Gamma(N(A+1)+n+j+1)$, where $1\leq j \leq \sum_{i=0}^\infty i k_i$. In the large $N$ limit those multipliers will always diminish any contribution that will be introduced by $\displaystyle \prod_{i=0}^\infty \left[\frac{C_{A+1+i}\Gamma(A+2+i)}{C_A\Gamma(A+1)}\right]^{k_i}\Big/\prod_{i=0}^\infty k_i!$. Eventually we are left with the expression
%%%%%%%%%%%%%%%%%%%%%%%%%%%%%%%%%%%%%%%%%%%%%%%
\begin{equation}
{{Q}}_t(n)\underset{N\to\infty}{\sim}
\displaystyle
\frac{\left(C_A\Gamma(A+1)t^{A+1}\right)^N}{\Gamma(N(A+1)+1)}
\sum_{n=0}^N
\frac{\left[t\frac{C_{A+1}\Gamma(A+2)}{C_A\Gamma(A+1)}\right]^{n}}{n!}
\frac{\left(N(A+1)\right)! N!}{(N-n)!\left(N(A+1)+n\right)!}.
    \label{qtnlong03}
\end{equation}
%%%%%%%%%%%%%%%%%%%%%%%%%%%%%%%%%%%%%%%%%%%%%%%
For the fraction of factorials we use $N!/(N-n)!=(-1)^n(-(N-n+1))(-(N-n+2))\dots (-N)$ and 
$(N(A+1)+n)!/(N(A+1))!=(N(A+1)+1)\dots(N(A+1)+n)$ to obtain
%%%%%%%%%%%%%%%%%%%%%%%%%%%%%%%%%%%%%%%%%%%%%%%%%
\begin{equation}
    {{Q}}_t(n)\underset{N\to\infty}{\longrightarrow}
\displaystyle
\frac{\left(C_A\Gamma(A+1)t^{A+1}\right)^N}{\Gamma(N(A+1)+1)}
\sum_{n=0}^N
\frac{\left[-t\frac{C_{A+1}(A+1)}{C_A}\right]^{n}}{n!}\frac{(-N)_n}{(N(A+1)+1)_n},
    \label{qtnshort01}
\end{equation}
%%%%%%%%%%%%%%%%%%%%%%%%%%%%%%%%%%%%%%%%%%%%%%%%%%%
where $(a)_b=\Gamma(a+b)/\Gamma(a)$ is the Pochhammer symbol~\cite{Abramowitz}. By using the fact that $(-N)_n=0$ for any $n>N$, Eq.~(\ref{qtnlong}) is rewritten as 
%%%%%%%%%%%%%%%%%%%%%%%%%%%%%%%%%%%%%%%%%%%%%%%%%
\begin{equation}
    {{Q}}_t(n)\underset{N\to\infty}{\longrightarrow}
\displaystyle
\frac{\left(C_A\Gamma(A+1)t^{A+1}\right)^N}{\Gamma(N(A+1)+1)}
{}_1F_1\left(-N;N(A+1)+1;-t\frac{C_{A+1}(A+1)}{C_A}\right),
    \label{qtnshort02}
\end{equation}
%%%%%%%%%%%%%%%%%%%%%%%%%%%%%%%%%%%%%%%%%%%%%%%%%%%
where ${}_1F_1(a;b;z)=\sum_{n=0}^\infty(a)_n z^n\Big/(b)_n n!$ is the Kummer's function of the first kind~\cite{Abramowitz}. The Kummer function ${}_1F_1(a;b;z)$ satisfies the second order differential equation 
$z\frac{d^2{}_1F_1(a;b;z)}{dz^2}+(b-z)\frac{d{}_1F_1(a;b;z)}{dz}-a{}_1F_1(a;b;z)=0$, that for the specific parameters of Eq.~(\ref{qtnshort02}) ($a=-N$ and $b=N(A+1)+1$) is given by 
%%%%%%%%%%%%%%%%%%%%%%%%%%%%%%%%%%%%%%%%%%%%%%%%%%%%%%%%%
\begin{equation}
    z\frac{d^2{}_1F_1}{dz^2}+(N(A+1)+1-z)\frac{d {}_1F_1}{dz}+N{}_1F_1=0.
    \label{kummerf}
\end{equation}
%%%%%%%%%%%%%%%%%%%%%%%%%%%%%%%%%%%%%%%%%%%%%%%%%%%%%%%%%%%
Multiplying Eq.~(\ref{kummerf}) by $1/N$ and taking the $N\to\infty$ limit we obtain $\frac{d {}_1F_1}{dz}+\frac{1}{A+1}{}_1F_1=0$ which provides the asymptotic behavior 
%%%%%%%%%%%%%%%%%%%%%%%%%%%%%%%%%%%%%%%%%%%%%%%%%%%%%%%%%
\begin{equation}
    {}_1F_1(-N;N(A+1)+1;z)\underset{N\to\infty}{\sim}
    \exp\left(-\frac{z}{A+1}\right).
    \label{kummersol}
\end{equation}
%%%%%%%%%%%%%%%%%%%%%%%%%%%%%%%%%%%%%%%%%%%%%%%%%%%%%%%%%
Eq.~(\ref{kummersol}) and Eq.~(\ref{qtnshort02}) give the asymptotic behavior of $Q_t(N)$
%%%%%%%%%%%%%%%%%%%%%%%%%%%%%%%%%%%%%%%%%%%%%%%%%%%%%%%%%
\begin{equation}
Q_t(N)\underset{N\to\infty}{\sim}
\displaystyle
\frac{\left(C_A\Gamma(A+1)t^{A+1}\right)^N}{\Gamma(N(A+1)+1)}
e^{\frac{C_{A+1}}{C_A}t}.
    \label{qtnassymptotic}
\end{equation}
%%%%%%%%%%%%%%%%%%%%%%%%%%%%%%%%%%%%%%%%%%%%%%%%%%%%%%%%%%
Finally, by using the Stirling's approximation we obtain
%%%%%%%%%%%%%%%%%%%%%%%%%%%%%%%%%%%%%%%%%%%%%%%%
\begin{equation}
    Q_t(N)\underset{N\to\infty}{\sim}
    e^{-N(A+1)\left( -\log(\frac{e(C_A\Gamma(A+1))^{\frac{1}{A+1}}}{A+1}\frac{t}{N})-\frac{C_{A+1}}{(A+1)C_A}\frac{t}{N}\right)}
%    \begin{cases}
 %   e^{\frac{C_{A+1}}{C_A}t}&  m= 1\\
  %  1              & m>1
%\end{cases}
    \label{qtnlimitApp}
\end{equation}
%%%%%%%%%%%%%%%%%%%%%%%%%%%%%%%%%%%%%%%%%%%%%%%%%%

\subsection{Details of $P(X,t)$ in Fig.~\ref{pxtexample}}
\label{appendixB}

The case of Gaussian $f(x)$ and uniform $\psi(\tau)$ produce $A=0$ and $C_A=1$, $C_{A+1}=0$.  In this case $K(N)=-X^2/2\delta^2N+N(\log(t)-\log(N)+1)$. The maxima is obtained for $N^*=(|X|/\delta)/W_0(X^2/t^2\delta^2)^{1/2}$ and $P(X,t)=\exp\left(K(N^*)\right)\big/\sqrt{2\pi K''(N^*)}$.
The case of uniform $f(x)$ and Dagum distribution $\psi(\tau)=1/(1+\tau)^2$ produce $A=0$ and $C_A=1$, $C_{A+1}=-2$. In this case of uniform $f(x)$, $\beta\to\infty$ and careful calculation of the limits  gives $K(N)=t(\frac{1}{N}\exp(N/2)-2)-N|X|$ and the maxima is obtained for $N^*=-2W_{-1}(-\frac{t}{4|X|})$. Notice that this time  $W_{-1}(y)$ is the lower branch of Lambert $W$ function, $W_{-1}(y)$ is defined for $-\frac{1}{e}\leq y < 0$. Finally, $P(X,t)=\exp\left(K(N^*)\right)\big/\sqrt{2\pi K''(N^*)}$.

\bibliographystyle{apsrev4-1} % Tell bibtex which bibliography style to use
\bibliography{./bibnonGaussian} % Tell bibtex which .bib file to use (this one is some example file in TexLive's file tree)

\begin{thebibliography}{50}
\bibitem [{lam()}]{lambertMath}%
  \BibitemOpen
  \href@noop {} {}\bibinfo {note} {This function is tabulated in
  Mathematica}\BibitemShut {NoStop}%
\end{thebibliography}

\end{document}